\documentclass[twoside,twocolumn,american,aps,superscriptaddress,showpacs,prl]{revtex4}
\usepackage[T1]{fontenc}
\usepackage[latin9]{inputenc}
\usepackage{color}
\usepackage{amsmath}
\usepackage{graphicx}
\usepackage{amssymb}
\usepackage{esint}

\makeatletter

\@ifundefined{textcolor}{}
{%
 \definecolor{BLACK}{gray}{0}
 \definecolor{WHITE}{gray}{1}
 \definecolor{RED}{rgb}{1,0,0}
 \definecolor{GREEN}{rgb}{0,1,0}
 \definecolor{BLUE}{rgb}{0,0,1}
 \definecolor{CYAN}{cmyk}{1,0,0,0}
 \definecolor{MAGENTA}{cmyk}{0,1,0,0}
 \definecolor{YELLOW}{cmyk}{0,0,1,0}
 }

\makeatother

\usepackage{babel}

\begin{document}

\title{Polarons and Molecules in a Two-Dimensional Fermi Gas }

\author{Sascha Zöllner}

\email{zoellner@nbi.dk}

\affiliation{The Niels Bohr International Academy, The Niels Bohr Institute, Blegdamsvej
17, 2100 Copenhagen, Denmark}

\author{G.\ M. Bruun}

\affiliation{Department of Physics and Astronomy, Aarhus University, 8000 Aarhus,
Denmark}

\author{C.\ J.\ Pethick}

\affiliation{The Niels Bohr International Academy, The Niels Bohr Institute, Blegdamsvej
17, 2100 Copenhagen, Denmark}

\affiliation{NORDITA, Roslagstullsbacken 23, 10691 Stockholm, Sweden}
\begin{abstract}
We study an impurity atom in a two-dimensional Fermi gas using variational
wave functions for (i) an impurity dressed by particle-hole excitations
(polaron) and (ii) a dimer consisting of the impurity and a majority
atom. In contrast to three dimensions, where similar calculations
predict a sharp transition to a dimer state with increasing interspecies
attraction, we show that the polaron ansatz always gives a lower energy.
However, the exact solution for a heavy impurity reveals that both
a two-body bound state and distortions of the Fermi sea are crucial.
This reflects the importance of particle-hole pairs in lower dimensions
and makes simple variational calculations unreliable. We show that
the energy of an impurity gives important information about its dressing
cloud, for which both ansätze give inaccurate results.

\end{abstract}


\date{November 16, 2010}

\maketitle
Since interactions between atoms can be tuned to essentially any value,
cold atomic gases provide a unique opportunity for studying experimentally
many-body physics in regimes that cannot be realized in other systems.
Recently, much attention has been given to the problem of a Fermi
gas with a low concentration of a second species, a so-called highly
imbalanced gas (see, e.g., \cite{schunck07,schirotzek09,nascimbene09}).
One fundamental problem is the nature of the ground state of a single
impurity atom in a Fermi gas. For weak interspecies attraction, the
ground-state energy is well described in terms of a state with an
impurity atom dressed by a single particle--hole excitation of the
Fermi sea, often referred to as a {}``polaron'' \cite{chevy06,combescot07},
while for strong attraction, a state based on a molecular picture
gives a lower energy \cite{prokofev08,punk09}. The transition between
the two states is predicted to be sharp \cite{prokofev08,bruun10}.

It is natural to ask whether this picture persists in lower dimensions.
This is of theoretical interest, since on general grounds one would
expect quantum fluctuations, in this case the creation of many particle--hole
pairs, to play an important role. In addition, the problem is on the
verge of being investigated experimentally with the use of optical
lattices \cite{martiyanov10}. In one dimension, a polaronic description
gives qualitative agreement with known exact results \cite{giraud09}\emph{.}
In this paper, we consider the case of two dimensions. We perform
simple variational calculations based on the polaron and molecule
pictures, and these predict that the polaronic state has the lower
energy for all interaction strengths, in marked contrast to what happens
in three dimensions. For an infinitely massive impurity, the problem
may be solved exactly, and the results show that the actual ground
state incorporates aspects of both pictures: A two-body bound state
is present for all coupling strengths, in addition to distortions
of the continuum states. We show that the energy of an impurity gives
important information about correlations in its vicinity and about
mutual interactions between impurities at nonzero density.

\paragraph*{Model\label{sec:Model}}

We consider a uniform two-dimensional (2D) Fermi gas of atoms of species
$a$, to which is added a single impurity atom of species $b$. The
two species may be either different hyperfine states of the same element
or different atomic species, in which case $b$ may be bosonic or
fermionic. The 2D confinement may be realized by a very tight trapping
potential in the transverse direction \cite{martiyanov10} and weak
longitudinal trapping. For densities low enough that only s-wave interactions
are important, the Hamiltonian reads \[
H\!=\!\sum_{\mathbf{k}}\epsilon_{\mathbf{k}}^{(a)}a_{\mathbf{k}}^{\dagger}a_{\mathbf{k}}+\!\sum_{\mathbf{k}}\epsilon_{\mathbf{k}}^{(b)}b_{\mathbf{k}}^{\dagger}b_{\mathbf{k}}+\!\sum_{\mathbf{kk}'\mathbf{q}}\!\frac{v(q)}{V}a_{\mathbf{k}+\mathbf{q}}^{\dagger}a_{\mathbf{k}}b_{\mathbf{k}'-\mathbf{q}}^{\dagger}b_{\mathbf{k}'},\]
 since $a$-$a$ interactions may be neglected because of the Pauli
principle. The single-particle eigenstates are $\langle\mathbf{x}|\mathbf{k}\rangle=e^{i\mathbf{k}\cdot\mathbf{x}}/\sqrt{V}$,
where we take the system to be enclosed in a 2D box of volume $V\equiv L^{2}$
with periodic boundary conditions. The single-particle energies are
$\epsilon_{\mathbf{k}}^{(\sigma)}=k^{2}/2m_{\sigma}$, where $m_{\sigma}$
is the mass of species $\sigma$ (we take $\hbar=1$ throughout);
$a_{\mathbf{k}}(b_{\mathbf{k}})$ annihilates an $a(b)$ atom in state
$|\mathbf{k}\rangle$. The interaction is modeled as $v(q)=g_{2}$
for particle momenta less than a cutoff value $\Lambda$ and zero
otherwise. The coupling $g_{2}$ and the cutoff may be eliminated
in favor of the two-body binding energy $\epsilon_{B}\ge0$ \cite{randeria90},
\begin{equation}
\frac{1}{g_{2}}=-\frac{1}{V}\sum_{|\mathbf{p}|<\Lambda}\frac{1}{\epsilon_{B}+\frac{\mathbf{p}^{2}}{2\mu}}=-\frac{2\mu}{4\pi}\ln\left(1+\frac{\Lambda^{2}/2\mu}{\epsilon_{B}}\right),\label{eq:renorm}\end{equation}
 with the reduced mass $\mu=(m_{a}^{-1}+m_{b}^{-1})^{-1}$. For $\epsilon_{B}\ll\Lambda^{2}/2\mu$,
none of the results depend on the cutoff, and we take $\Lambda\to\infty$
at the end of the calculation.

\paragraph{Polaron}

We describe the state using the variational ansatz \cite{chevy06}
\begin{equation}
\Psi_{P}=\left(\phi_{0}b_{\mathbf{0}}^{\dagger}+\negthickspace\sum_{|\mathbf{q}|<k_{F}<|\mathbf{k}|}\negthickspace\phi_{\mathbf{kq}}\, b_{\mathbf{q-k}}^{\dagger}a_{\mathbf{k}}^{\dagger}a_{\mathbf{q}}\right)|N\rangle,\label{eq:Psi_Chevy}\end{equation}
 which describes an impurity atom $b$ (here at zero momentum) dressed
by a cloud of particle-hole excitations of the ground state $|N\rangle$
of $N$ $a$-atoms. This state is an expansion of the true state in
terms of numbers of particle-hole excitations, truncated at first
order. Minimizing the energy functional for this ansatz leads to equations
for the energy of the state relative to the ground-state energy of
$N$ $a$-atoms, $E\equiv\langle H\rangle_{\Psi_{P}}-E_{0}^{(N)}$,
and the amplitudes $\phi_{\mathbf{kq}}$: \begin{equation}
E=\frac{1}{V}\!\sum_{\mathbf{q}}n_{q}T_{\mathbf{q}}(E+\epsilon_{\mathbf{q}}^{(a)}),\;\frac{\phi_{\mathbf{kq}}}{\phi_{0}}=\frac{1}{V}\frac{T_{\mathbf{q}}(E+\epsilon_{\mathbf{q}}^{(a)})}{E-\Delta\epsilon_{\mathbf{kq}}}.\label{eq:E_Chevy-1}\end{equation}
 Here the T-matrix $T_{\mathbf{q}}(E+\epsilon_{\mathbf{q}}^{(a)})^{-1}=g_{2}^{-1}-V^{-1}\sum_{|\mathbf{k}|<\Lambda}(1-n_{k})\left(E-\Delta\epsilon_{\mathbf{kq}}\right)^{-1}$
plays the role of an effective interaction, with $\Delta\epsilon_{\mathbf{kq}}\equiv\epsilon_{\mathbf{k}}^{(a)}-\epsilon_{\mathbf{q}}^{(a)}+\epsilon_{\mathbf{q-k}}^{(b)}$,
and $n_{q}\equiv\Theta(k_{F}-q)$ is the Fermi function, $k_{F}$
being the Fermi momentum. In the thermodynamic limit (keeping the
majority density $n_{a}=N/V=k_{F}^{2}/4\pi$ fixed), this becomes
an integral equation, whose root $E$ may be found numerically.

\begin{figure}
\begin{centering}
\includegraphics[width=0.8\columnwidth]{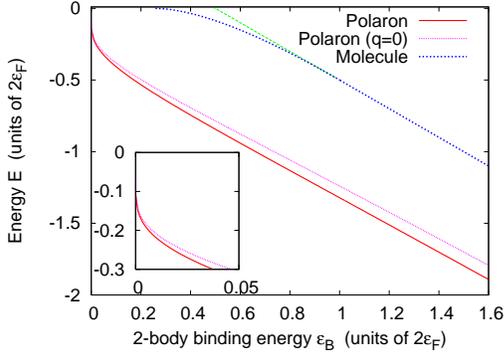} 
\par\end{centering}

\caption{(color online) Energy $E$ as a function of $\epsilon_{B}$ for the
polaron ansatz (for the $q=0$ approximation, see text) and the dimer
ansatz. (For $\epsilon_{B}<2\epsilon_{F}$, the dimer has momentum
$p_{M}\neq0$; the thin straight line indicates the dimer energy at
$p_{M}=0$.) \label{fig:E_2D}}

\end{figure}

For now, we will focus on the equal-mass case, $m_{\sigma}\equiv m=2\mu$.
The energy $E(\epsilon_{B})$ obtained numerically is shown in Fig.~\ref{fig:E_2D}.
In the \emph{weak-coupling }limit\emph{ }$\epsilon_{B}\ll\epsilon_{F}\equiv k_{F}^{2}/2m$,
the energy $E\simeq-2\epsilon_{F}/\ln(2\epsilon_{F}/\epsilon_{B})$
decreases very rapidly with increasing $\epsilon_{B}$. This result
can be interpreted as a mean-field shift $E\sim n_{a}\bar{g},$ where
the density-dependent effective interaction $\bar{g}=-4\pi/m\ln(n_{a}4\pi\hbar^{2}/m\epsilon_{B})$
is the leading term in $T_{\mathbf{q}}$ for $E,\epsilon_{B}\to0$.
In the \emph{strong-coupling} limit $\epsilon_{B}\gg\epsilon_{F}$,
we find $E\simeq-\epsilon_{B}-\eta\epsilon_{F}$, with $\eta\approx0.26$.
Some analytical insight into these results may be obtained by approximating
$\Delta\epsilon_{\mathbf{kq}}=(\mathbf{k}^{2}-\mathbf{k}\cdot\mathbf{q})/m$
by $\mathbf{k}^{2}/m$, as has proven qualitatively correct in the
3D case \cite{combescot08}. This {}``$q=0$'' approximation gives
the equation $E\approx n_{a}T_{\mathbf{0}}(E)=-\epsilon_{B}\exp(2\epsilon_{F}/|E|)+2\epsilon_{F}$,
whose solution is plotted for comparison in Fig.~\ref{fig:E_2D}.
It gives the correct weak-coupling limit; in the strong-coupling limit
it yields $E\simeq-\epsilon_{B}-2\epsilon_{F}^{2}/\epsilon_{B}$,
which misses the term $O(\epsilon_{F})$. We have also studied the
case of arbitrary $m_{b}$ and find that the mass dependence of the
polaron energy is weak if energies are expressed in units of $k_{F}^{2}/2\mu$.

At first sight, it is somewhat surprising that for strong coupling
the polaronic ansatz leads to an energy close to that of a molecule
in vacuum. In this limit, the overlap probability of $\Psi_{P}$ with
the non-interacting ground state is given by $Z\equiv|\phi_{0}|^{2}\simeq2\epsilon_{F}/\epsilon_{B}$,
which tends to zero. Thus the state is comprised mainly of holes and
an $(ab)$ dimer, as may be seen from Eq.\ (\ref{eq:Psi_Chevy}),
and the coefficients $\phi_{\mathbf{kq}}$ {[}Eq.\ (\ref{eq:E_Chevy-1}){]}
reduce to the wave function for a molecule in vacuo plus a hole. The
leading contribution to the energy is thus the energy of a molecule
in free space, since the hole has an energy of at most $\epsilon_{F}$.

\paragraph*{Molecule}

In three dimensions, it has been demonstrated that a simple variational
wave function based on a molecular picture gives a lower energy than
the polaron ansatz (\ref{eq:Psi_Chevy}) for sufficiently strong attraction
\cite{punk09}, and we investigate whether this happens in two dimensions.
In its simplest form, such an $ab$ dimer with zero total momentum
may be modeled using the trial state \cite{punk09} \begin{equation}
\Psi_{M}=\sum_{|\mathbf{k}|>k_{F}}\varphi_{\mathbf{k}}b_{-\mathbf{k}}^{\dagger}a_{\mathbf{k}}^{\dagger}|N-1\rangle,\label{eq:Psi_Mol}\end{equation}
 which corresponds to a correlated $ab$ pair in states with momentum
greater than $k_{F}$ and a Fermi sea with $N-1$ noninteracting atoms.
Minimizing the energy functional leads to the following equation for
the energy $E_{M}$ (again relative to that of the $N$-particle Fermi
sea), \begin{equation}
\frac{1}{g_{2}}=-\frac{1}{V}\negmedspace\sum_{k_{F}<|\mathbf{k}|<\Lambda}\frac{1}{\frac{k^{2}}{2\mu}-(E_{M}+\epsilon_{F})},\label{eq:E2D_Mol}\end{equation}
 whose solution is\[
E_{M}=-\epsilon_{B}+\frac{k_{F}^{2}}{2m_{b}},\]
 and $\varphi_{\mathbf{k}}\propto\Theta(k-k_{F})/(\epsilon_{\mathbf{k}}^{(b)}-E_{M})$.
This result is simple because in 2D the density of states is independent
of energy: The energy of the dimer is shifted by the kinetic energy
of the lowest state not Pauli-blocked.  In this approximation, the
bare zero-momentum dimer state is always energetically less favorable
than the polaronic solution, and there is no sharp transition (Fig.~\ref{fig:E_2D}),
in contrast to the 3D case.

 The question arises whether dimers with nonzero momentum have a
lower energy. The extension of (\ref{eq:Psi_Mol}) to describe a dimer
with momentum $\mathbf{p}$ is $\Psi_{M}^{(\mathbf{p})}=\sum_{|\mathbf{k}|>k_{F}}\varphi_{\mathbf{k}}^{(\mathbf{p})}a_{\mathbf{k}}^{\dagger}b_{\mathbf{p-k}}^{\dagger}|N-1\rangle$,
which leads to an equation similar to Eq.\ (\ref{eq:E2D_Mol}). The
solution is \begin{equation}
E_{M}(p)-E_{M}(0)=\frac{p^{2}}{2M}-\frac{k_{F}^{2}}{2\mu}\left[1+\left(\frac{M/m_{b}}{pa_{2}}\right)^{2}\right]^{-1}\negthickspace,\end{equation}
 where $\epsilon_{B}\equiv\hbar^{2}/2\mu a_{2}^{2}$ defines the 2D
scattering length $a_{2}$. For $\epsilon_{B}>\frac{k_{F}^{2}}{2\mu}\frac{m_{a}}{m_{b}}$,
the dimer energy is minimal for $p=0$. For $\epsilon_{B}<\frac{k_{F}^{2}}{2\mu}\frac{m_{a}}{m_{b}}$,
though, a dimer at $p=p_{M}\neq0$ is favorable, because the kinetic-energy
increase is outweighed by the reduced Pauli-blocking shift %
\footnote{Although the lowest-energy dimer has nonzero momentum, its velocity
vanishes by definition, $\partial_{p}E(p_{M})=0$.%
}. For concreteness, consider the equal-mass case: The dimer momentum
$p_{M}=2\sqrt{k_{F}a_{2}-1}/a_{2}$ changes smoothly from $p_{M}=0$
(at $\epsilon_{B}=2\epsilon_{F}$) to $p_{M}\to k_{F}$ as $\epsilon_{B}\to\epsilon_{F}/2$,
where the energy reaches the continuum threshold $E_{M}=0$ (Fig.~\ref{fig:E_2D}).
In that regime, 
$E_{M}=-\epsilon_{F}+2\sqrt{\epsilon_{B}}(\sqrt{2\epsilon_{F}}-\sqrt{\epsilon_{B}})$.
Thus, the bare dimer is energetically unfavorable compared with the
polaronic ansatz for any coupling strength. To obtain more insight,
we now examine an exact solution.

\paragraph*{Infinite-mass limit}

We consider the case where the impurity atom is very massive, $m_{b}\gg m_{a}$.
In addition to being exactly soluble by use of Fumi's theorem \cite{mahan},
this is also relevant to experiments with atomic mixtures (e.g., Li
and Yb \cite{okano10}). The massive impurity may be treated as a
static defect, which has two effects on the Fermi gas: It creates
a two-body bound state with energy $-\epsilon_{B}$, and it also shifts
the energy levels in the continuum. In any number of dimensions, the
energy shift of an s-wave level in the continuum is given by $-k\delta_{0}(k)/m_{a}R$,
where $\delta_{0}(k)$ is the s-wave phase shift and $R$ the radius
of the sphere containing the Fermi gas. The density of s-wave levels
per unit interval in $k$ is $\pi/R$. The total energy change when
an impurity is added to the $N$-atom system is \begin{equation}
E=-\epsilon_{B}-\frac{1}{m_{a}\pi}\int_{0}^{k_{F}}\negthickspace dk\, k\delta_{0}(k).\label{Fumi}\end{equation}
 In 2D, the scattering phase shift is given by $\cot\delta_{0}(k)=\ln(k^{2}/2m_{a}\epsilon_{B})/\pi$
in the zero-range limit \cite{randeria90}. 

For large $\epsilon_{B}$, $\Delta E\equiv E+\epsilon_{B}\simeq-\epsilon_{F}+\frac{\epsilon_{F}}{\ln(\epsilon_{B}/\epsilon_{F})}$.
The leading term corresponds to the fact that the phase shift is close
to $\pi$, and consequently each of the continuum s-levels has become
close to the next lower one in the absence of the impurity, thereby
lowering the energy by $\epsilon_{F}$. The logarithmic term may be
thought of as an effective atom-dimer repulsion. This is in stark
contrast to the result $\Delta E_{M}=k_{F}^{2}/2m_{b}\to0^{+}$ ($m_{b}\to\infty$)
obtained from the dimer ansatz. Moreover, for the polaron in the heavy-impurity
limit, we find numerically 
$\Delta E\approx-0.14\epsilon_{F}+O(n_{a}^{2})$, which is much smaller
in magnitude than the exact result, $-\epsilon_{F}$.

The exact result shows that the ground state has a bound two-body
state and that levels in the continuum are modified. It therefore
incorporates aspects of both the polaron and the dimer pictures. The
missing ingredient in the trial states we have used is components
with higher number of particle-hole pairs: This is treated only to
first (zeroth) order in the polaron (dimer) ansatz. In the language
of perturbation theory, the T-matrix employed in the polaron ansatz
is calculated in the ladder approximation, and thus contains only
particle--particle and hole--hole scattering. However, it is known
that the T-matrix for scattering of a fermion from a massive impurity
is essentially independent of the presence of the medium, because
Pauli blocking of ladder diagrams is compensated by impurity--hole
scattering \cite{bruun10a}. Our calculations show that the higher-order
impurity--hole scattering processes change $\Delta E$ to leading
order from $-0.14\epsilon_{F}$ (for one particle-hole pair) to $-\epsilon_{F}$
in the exact result.

However, the bare-dimer picture worked reasonably well in 3D, predicting
a molecular transition in agreement with Monte-Carlo results. To understand
this paradox, let us look at the role of dimensionality in the dimer
problem. Solving (\ref{eq:E2D_Mol}) in $D=1,\dots,3$ (for equal
masses) yields an energy of the form $E_{M}\simeq-\epsilon_{B}-\epsilon_{F}+c_{D}\epsilon_{F}(\frac{2\epsilon_{F}}{\epsilon_{B}})^{(D-2)/2}$
as $\epsilon_{B}\to\infty$. The last term corresponds to an upshift
of the dimer energy due to Pauli blocking of the states $|\mathbf{q}|<k_{F}$.
In $D=3$ the upshift vanishes in the limit $\epsilon_{B}\gg2\epsilon_{F}$:
This is because the density of states $\varrho(\epsilon)\propto\sqrt{\epsilon}$
vanishes at low energies, so that the contribution from Pauli blocking
for $\epsilon<2\epsilon_{F}$ has a negligible weight for $\epsilon_{B}\to\infty$.
The situation is dramatically different in lower dimensions. In $D=2$,
the density of states is constant and thus leads to an interaction-independent
displacement of the vacuum energy by $2\times\epsilon_{F}$, recovering
the total shift $+\epsilon_{F}$. This also illuminates why bare dimers
should be even less favorable in 1D \cite{leskinen10}, where the
low-lying states have an even stronger weight $\varrho(\epsilon)\propto1/\sqrt{\epsilon}$,
leading to a diverging upshift for strong coupling.



\paragraph*{Dressing cloud of an impurity}

\begin{figure}
\begin{centering}
\includegraphics[width=0.7\columnwidth]{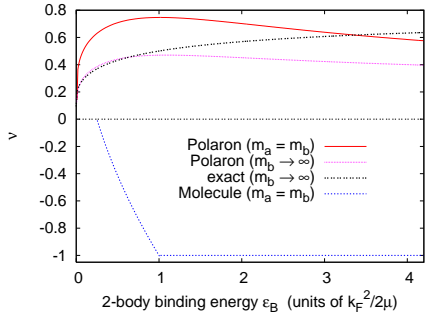} 
\par\end{centering}

\caption{(color online) Number of majority atoms \emph{$\nu$} in the dressing
cloud of an impurity\emph{ }as a function of $\epsilon_{B}$\emph{.
} \label{fig:induced} }

\end{figure}

Important information about the structure of the dressing cloud of
an impurity may be extracted from the results for the energy. As is
done in the theory of dilute mixtures of helium isotopes \cite{bardeen67},
we define the quantity $\nu=(\partial n_{a}/\partial n_{b})_{\mu_{a}}=-(\partial\mu_{b}/\partial n_{a})/(\partial\mu_{a}/\partial n_{a})$,
where $\mu_{\sigma}$ is the chemical potential of a $\sigma$-atom.
Physically, this is the number of $a$-atoms in the dressing cloud
of an impurity. The requirement that $\mu_{a}$ be held fixed ensures
that far from the impurity, the density of $a$-atoms is unchanged
by addition of the impurity. For $n_{b}\ll n_{a}$, this number can
be deduced from the single-impurity energy, $\nu=-\partial E/\partial\epsilon_{F}$,
which is plotted in Fig.~\ref{fig:induced}. As expected, $\nu$
tends to zero in the weak-coupling limit, $\nu\simeq2/\ln(2\epsilon_{F}/\epsilon_{B})$,
for $m_{a}=m_{b}$. For $m_{b}\to\infty$, we can infer that there
is exactly one dressing atom as $\epsilon_{B}\to\infty$, $\nu\to1$.
This contrasts with the polaron ansatz, which for $m_{b}\to\infty$
predicts $\nu_{P}\to\eta\approx0.14$ (following a peak near $\epsilon_{B}=2\epsilon_{F}$),
illustrating that the single-particle--hole picture highly underestimates
the impurity dressing. For comparison, the bare-dimer ansatz predicts
the unphysical result $\nu_{M}=-1$, amounting to a deficit of atoms
in the dressing cloud due to Pauli blocking.

\paragraph*{Nonzero impurity density}

An intriguing question concerns the behavior at nonzero impurity density:
Do the {}``dressed'' impurity atoms behave as fermions, bosons,
or neither of them? For weak attraction, it is not implausible that
the dressed impurities have the same quantum statistics as bare ones.
By contrast, for strong attraction, one may expect the basic degrees
of freedom to be best described in terms of $ab$ dimers, which are
bosons for fermionic impurities and vice versa. On the basis of simple
arguments, we cannot arrive at a definite conclusion about the statistics
obeyed by the elementary excitations; to do so, it would be necessary
to investigate the importance of exchange processes in a system with
two impurities %
\footnote{Even if the polaron ansatz gives the lower energy, one cannot conclude
that the elementary excitations obey the same statistics as the impurity
atom, since the polaron state is dominated by a rather incoherent
superposition of many dimer--hole configurations.%
}.  

Let us consider the case when the quasiparticles are fermionic. This
could apply for weakly interacting fermionic impurities, but also
for bosonic ones if they form a tightly bound dimer with a majority
atom.  We now show that the single-impurity findings have implications
for the thermodynamic properties at nonzero concentration $n_{b}/n_{a}\ll1$.
The total energy density $\mathcal{E}$ of such a Fermi liquid then
reads $\mathcal{E}(n_{b})\simeq\mathcal{E}(0)+\mathcal{E}_{0}(n_{b})+E(\epsilon_{B})n_{b}+\frac{1}{2}fn_{b}^{2}$,
where\emph{ $\mathcal{E}(0)$ }is the majority energy, $\mathcal{E}_{0}(n_{b})=\pi n_{b}^{2}/m_{b}^{*}$
denotes the kinetic-energy density, with the effective mass $m_{b}^{\star}$
modified by interactions, and the term $n_{b}E(\epsilon_{B})$ gives
the energy reduction due to binding of independent quasiparticles.
Even in the absence of \emph{direct }interactions between \emph{b}-fermions,
there is an \emph{induced }interaction between them, mediated by the
majority Fermi gas \cite{yu10}. It turns out to be repulsive owing
to the Pauli principle and is characterized by the Landau parameter
$f=\nu{}^{2}\partial\epsilon_{F}/\partial n_{a}$. Note that, since
in 2D the density of states $\partial n_{a}/\partial\epsilon_{F}=m_{a}/2\pi$
is constant, $f$ is nonzero for $n_{a}\to0$. We mention that for
bosonic quasiparticles, the effective interaction follows in a similar
fashion \cite{viverit00}, the difference being that there is a direct
s-wave interaction and that the induced interaction is attractive.
However, how that influences the induced interactions depends nontrivially
on the degeneracy of the bosons and is left for future studies.

Finally, we mention that the properties of impurities can be probed
using techniques similar to those in 3D. By exciting collective oscillations,
the effective mass is accessible \cite{nascimbene09}. With increasing
coupling, this tends to $\infty$ for the polaron and to $M$ for
the dimer ansatz. Another key tool is radio-frequency spectroscopy
\cite{schirotzek09}, where a $b$ atom is transferred from its initial
hyperfine level to an empty one via a pulse of frequency $\omega$.
For a polaron, the transition rate \emph{$\Gamma(\omega)$} decomposes
into a quasiparticle peak $\propto Z\delta(\omega-|E|)$, indicating
the polaron contribution, and an incoherent background $\Gamma_{\mathrm{inc}}(\omega)$,
which increases as $(\omega-|E|)^{3/2}$ for $0\le\omega-|E|\ll\epsilon_{F}$
and falls off as $\omega^{-2}$ for $\omega-|E|\gg\epsilon_{F}$,
if final-state interactions are ignored. This contrasts with the dimer
ansatz, which yields $\Gamma_{M}(\omega)\propto\Theta(\omega-|E|)/\omega^{2}$
without any quasiparticle peak.

In summary, using variational wave functions, we find no evidence
for a sharp transition between the polaron and the molecular picture
in two dimensions. Comparison with the exact result for a heavy impurity
shows that both ansätze lead to inaccurate results for the dressing
cloud of the impurity in the strong-coupling limit. This reveals the
key role of many particle-hole pairs, and it reflects the importance
of quantum fluctuations in lower dimensions. We conclude that more
work is needed to understand the nature of the ground state of an
impurity with finite mass in a two-dimensional Fermi gas. 

We are grateful to H.~Smith, Z.~Yu, and P.~Massignan for helpful
discussions. SZ was supported by the German Academy of Sciences Leopoldina
(LPDS 2009-11).

\end{document}